\documentstyle[12pt,epsfig]{article}
\newcommand{\be}{\begin{equation}}
\newcommand{\ee}{\end{equation}}
\newcommand{\bea}{\begin{eqnarray}}
\newcommand{\eea}{\end{eqnarray}}

\newcommand{\pup}{p^\uparrow}

\newcommand{\Lup}{\Lambda^\uparrow}

\newcommand{\nd}{\noindent}
\def\lsim{\mathrel{\rlap{\lower4pt\hbox{\hskip1pt$\sim$}}\raise1pt\hbox{$<$}}}
\def\gsim{\mathrel{\rlap{\lower4pt\hbox{\hskip1pt$\sim$}}\raise1pt\hbox{$>$}}}
\newcommand{\NP}[1]{{\it Nucl.\ Phys.}\ {\bf #1}}
\newcommand{\ZP}[1]{{\it Z.\ Phys.}\ {\bf #1}}
\newcommand{\PL}[1]{{\it Phys.\ Lett.}\ {\bf #1}}
\newcommand{\PR}[1]{{\it Phys.\ Rev.}\ {\bf #1}}
\newcommand{\PRL}[1]{{\it Phys.\ Rev.\ Lett.}\ {\bf #1}}

\begin{document}
%%%%%%%%%%%%%%%%%%%%%%%%%%%%%%%%%%%%%%%%%%%%%%%%%%%%%%%%%%%%%%%%%%%%%%%%%%%%%%
\begin{flushright} 
%DFTT 28/2001 \\ 
%INFNCA-TH0208 \\ 
%hep-ph/0403114 \\ 
\end{flushright} 
\vskip 1.5cm
\begin{center}
{\bf Accessing transversity via $J/\psi$ production in polarized 
{\mbox{\boldmath $\pup \!\! \bar p^{\,\uparrow}$}} interactions}\\
\vskip 0.8cm
{\sf M.~Anselmino$^1$, V.~Barone$^2$, A.~Drago$^3$, N.N. Nikolaev$^{4,5}$}
\vskip 0.5cm
{\it $^1$ Dipartimento di Fisica Teorica, Universit\`a di Torino and \\
          INFN, Sezione di Torino, Via P. Giuria 1, I-10125 Torino, Italy}\\
\vspace{0.3cm}
{\it $^2$ Di.S.T.A., Universit\`a del Piemonte Orientale ``A. Avogadro'', \\
and INFN, Gruppo Coll. di Alessandria, 15100 Alessandria, Italy}\\
\vspace{0.3cm}
{\it $^3$  Dipartimento di Fisica, Universit\`a di Ferrara and\\
INFN, Sezione di Ferrara, 44100 Ferrara, Italy}\\
\vspace{0.3cm}
{\it $^4$ Institut f\"ur Kernphysik, Forschungszentrum J\"ulich, D-52425
J\"ulich, Germany}\\
\vspace{0.3cm}
{\it $^5$ L.D. Landau Institute for Theoretical Physics, 142432
Chernogolovka, Russia}\\
\end{center}

\vspace{1.5cm}

\begin{abstract}
We discuss the possibility of a direct access to transversity distributions
by measuring the double transverse spin asymmetry $A_{TT}$ in 
$\pup \, \bar p^{\,\uparrow} \to J/\psi \, X \to \ell^- \, \ell^+ \, X$ 
processes at future GSI-HESR experiments with polarized protons and 
anti-protons. In the $J/\psi$ resonance production region, with 
$30 \lsim s \lsim 45$ GeV$^2$, both the cross-section and $A_{TT}$ are 
expected to be sufficiently large to allow a measurement of $h_1^q(x,M^2)$; 
numerical estimates are given. 
%Also the channel 
%$\pup \, \bar p^{\,\uparrow} \to D \, X$ is briefly considered.
\vspace{0.6cm}

\end{abstract}
%{}~~~PACS numbers: %
%%%%%%%%%%%%%%%%%%%%%%%%%%%%%%%%%%%%%%%%%%%%%%%%%%%%%%%%%%%%%%%%%%%%%%%%%%%%%% 
\newpage 
%%%%%%%%%%%%%%%%%%%%%%%%%%%%%%%%%%%%%%%%%%%%%%%%%%%%%%%%%%%%%%%%%%%%%%%%%%%%%% 
\pagestyle{plain} 
\setcounter{page}{1} 
\nd 
{\bf 1. Introduction} 
\vskip 6pt 

Transversity is the last leading-twist missing information on the quark spin 
structure of the nucleon \cite{bdr}; 
whereas the unpolarized quark distributions, 
$q(x,Q^2)$, are well known, and more and more information is becoming available
on the helicity distributions $\Delta q(x,Q^2)$, nothing is experimentally 
known on the nucleon transversity distribution $h_1^q(x,Q^2)$ [also denoted by 
$\Delta_T q(x,Q^2)$ or $\delta q(x,Q^2)$]. From the theoretical side, there 
exist only a few and rather preliminary models for $h^q_1$. 
The reason why $h^q_1$, despite its fundamental importance, has never been 
measured is that it is a chiral-odd function, and consequently it decouples
from inclusive Deep Inelastic Scattering (DIS), which is our usual main source 
of information on the nucleon partonic structure. Since electroweak and strong 
interactions conserve chirality, $h^q_1$ cannot occur alone, but has to be 
coupled to a second chiral-odd quantity. 

This is possible, for example, in polarized Drell-Yan processes
\cite{h1}, where one measures the product of two transversity distributions, 
and in semi-inclusive DIS, where one couples $h^q_1$ to a new unknown,
chiral-odd, fragmentation function, the so-called Collins functions 
\cite{col}. Similarly, one could couple $h^q_1$ and the Collins function in 
transverse single spin asymmetries in inclusive processes like 
$\pup \, p \to \pi \, X$; or, one could couple $h^q_1$ to another polarized 
fragmentation function by studying the spin transfer in processes like 
$\ell \, \pup \to \ell \, \Lup \, X$ \cite{hil}.

HERMES collaboration have measured single spin asymmetries in semi-inclusive 
DIS processes \cite{herm} and, together with COMPASS experiment \cite{comp}, 
are still gathering data which should yield information on some combination 
of $h^q_1$ and the Collins function. However, it will not be easy to extract 
information on $h^q_1$ alone: the measured spin asymmetries can originate
also from the Sivers function \cite{siv}, a chiral-even spin property of 
quark distribution, rather than fragmentation; also, higher twist effects 
might still be sizeable at the $Q^2$ of the two experiments, thus making 
the interpretation of data less direct.

Measurement of transversity is planned at RHIC, in Drell-Yan processes
with transversely polarized protons, $p^\uparrow p^\uparrow \to
\ell^- \ell^+ \, X$, via the measurement of the double spin asymmetry:
\begin{equation}
A_{TT}^{pp} \equiv \frac{d\sigma^{\uparrow\uparrow} - 
d\sigma^{\uparrow\downarrow}} {d\sigma^{\uparrow\uparrow} 
+ d\sigma^{\uparrow\downarrow}},  
\end{equation}
which reads at leading order
\begin{equation}
A_{TT}^{pp} = \hat a_{_{TT}} \> 
\frac{\sum_q e_q^2 \left[ h_1^q(x_1, M^2) \, h_1^{\bar q}(x_2, M^2)
 +  h_1^{\bar q}(x_1, M^2) \, h_1^q(x_2, M^2) \right]}
{\sum_q e_q^2 \left[ q(x_1, M^2) \, \bar q(x_2, M^2)
 + \bar q(x_1, M^2) \, q(x_2, M^2) \right]}\>, \label{att}
\end{equation}
where $q = u, d, s$; $M$ is the invariant mass of the lepton 
pair and $\hat a_{TT}$ is the double spin asymmetry of the QED elementary
process, $q \, \bar q \to \ell^- \ell^+$ [see below, Eq. (\ref{atttp})].
In this case one measures the product of two transversity 
distributions, one for a quark and one for an anti-quark.
The latter (in a proton) is expected to be small; moreover, the QCD evolution 
of transversity is such that, in the kinematical regions of RHIC data, 
$h^q_1(x, Q^2)$ is much smaller than the corresponding values of 
$\Delta q(x,Q^2)$ and $q(x,Q^2)$. All this makes the Drell-Yan double spin 
asymmetry $A_{TT}$ measurable at RHIC very small, no more than a few percents 
\cite{bcd,mssv}. This would remain true at RHIC energies
even if polarized anti-protons were available \cite{bcd}.

One could consider the double spin asymmetry $A_{TT}$ also for other 
processes, like $\pup \pup \to jet + X$, $\pup \pup \to \gamma \, X$,
{\it etc.}; however, $A_{TT}$ always turns out to be very small 
\cite{ssv,msv}, so that accessing transversity at RHIC appears as a very
difficult task. 

The single spin asymmetries experimentally observed in 
$p^\uparrow p \to \pi \, X$ and $\bar p^{\,\uparrow} p \to \pi \, X$ processes 
\cite{ags,e704,star} can be interpreted in terms of transversity and Collins 
functions \cite{noic}; however, also contributions from Sivers function 
(with no transversity) are important \cite{nois} and these processes could 
hardly be used for extracting information on $h_1^q$ alone.   

Definite and direct information on transversity should be best obtained 
in processes and in kinematical regions such that: $h^q_1(x, Q^2)$ is sizeable,
it couples to itself rather than to other unknown quantities, and the related
physical observables do not receive large contributions from gluons (which 
do not carry any transversity). We discuss here such an ideal situation,
considering the possibility -- at the moment only at the stage of a proposal --
of having polarized anti-protons colliding on polarized protons in the 
High Energy Storage Ring at GSI \cite{pax}. 

In the next two Sections we shall then discuss lepton pair production in 
$\pup \, \bar p^{\,\uparrow}$ interactions in the following kinematical 
region:
\be
(30 \lsim s \lsim 45) \> {\rm GeV^2};
\quad\quad\quad
M \gsim 2 \> {\rm GeV}/c^2;
\quad\quad\quad
\tau = x_1x_2 = \frac{M^2}{s} \gsim 0.1 \>. \label{reg}
\ee 
In Section 4 we present some conclusions.  

\vspace{18pt}
\goodbreak
\nd
{\bf 2. {\mbox{\boldmath $A_{TT}$}} for Drell-Yan processes in 
{\mbox{\boldmath $\pup \!\! \bar p^{\,\uparrow}$}} interactions}
\nobreak
\vspace{6pt}
\nobreak

The unpolarized Drell-Yan cross-section in $p \, \bar p$ interactions is 
given, at LO, by:
\be
\frac{d\sigma}{d\Omega \, dx_1 \, dx_2} = 
\sum_q e_q^2 \left[ q(x_1, M^2) \, q(x_2, M^2)
 + \bar q(x_1, M^2) \, \bar q(x_2, M^2) \right] 
\frac{d\hat{\sigma}}{d\Omega} 
\label{unppp}
\ee
where 
\be
\frac{d\hat{\sigma}}{d\Omega} = \frac{\alpha^2}{12M^2}(1 + \cos^2\theta)
\label{unpqq}
\ee
is the cross-section for the elementary process $q\,\bar q \to \ell^-\ell^+$;
$\theta$ is the production angle in the rest frame of the lepton pair
(we follow the notations and geometrical configurations of Ref. \cite{bdr}) 
and $M$ is the invariant mass of the lepton pair.
In Eq. (\ref{unppp}) $x_1$ and $x_2$ are the usual momentum fractions 
carried by the (anti)quarks and all quark distributions {\it refer to protons} 
(a $\bar q$ distribution inside a $\bar p$ is the same as a $q$ inside a $p$, 
{\it etc.}). 

Usually, one integrates over all production angles of the lepton pair 
and uses, rather than the variables $x_1$ and $x_2$, other physical 
observables like $M$, $\tau$, $y$ or $x_F$, related to $x_1$, $x_2$ by:  
\bea
M^2 = x_1\,x_2\,s \equiv \tau\,s  
\quad\quad y \equiv \frac{1}{2} \, \ln \frac{x_1}{x_2} 
\quad\quad x_F \equiv 2\,q_L/\sqrt s = x_1 - x_2 \label{var1}\\ 
x_1 = \frac{\sqrt{x_F^2 + 4\tau} + x_F}{2} = \sqrt{\tau}\,e^y \quad\quad
x_2 = \frac{\sqrt{x_F^2 + 4\tau} - x_F}{2} = \sqrt{\tau}\,e^{-y} \>,
\label{var2}
\eea
where $q_L$ is the longitudinal momentum of the lepton pair and 
$\sqrt s$ is the total $p\,\bar p$ c.m. energy.

Using some of these variables Eq. (\ref{unppp}) can be written, for example, 
as 
\be
\frac{d\sigma}{dM^2\, dx_F} = 
\frac{4 \pi \alpha^2}{9 \, M^2 s \, (x_1 + x_2)}
\sum_q e_q^2 \left[ q(x_1, M^2) \, q(x_2, M^2)
 + \bar q(x_1, M^2) \, \bar q(x_2, M^2) \right] 
\label{unppp1}
\ee
with $x_1, x_2$ as given in Eq. (\ref{var2}) and $x_1x_2 = \tau = M^2/s$. 

In case of transversely polarized $p$ and $\bar p$ -- therefore transversely
polarized $q$ and $\bar q$ -- the elementary cross-section depends also 
on the azimuthal angle $\varphi = \Phi - \Phi_S$, that is the difference 
between the azimuthal angles of the lepton pair and the proton polarization:
\be
\frac12 \left[
\frac{d\hat{\sigma}^{\uparrow\uparrow}}{d\Omega} -
\frac{d\hat{\sigma}^{\uparrow\downarrow}}{d\Omega} \right]
\equiv \frac{d\Delta\hat{\sigma}}{d\Omega} =
\frac{\alpha^2}{12M^2} \, \sin^2\theta \cos(2\varphi) \>. \label{polqq} 
\ee

For the cross-section difference
\be 
%\sigma = \frac12 \, \left[ \sigma^{\uparrow\uparrow} +
%\sigma^{\uparrow\downarrow} \right] \quad\quad\quad  
\Delta\sigma = \frac12 \, \left[ \sigma^{\uparrow\uparrow} -
\sigma^{\uparrow\downarrow} \right] \>, \label{ds}
\ee
we have  
\be
\frac{d\Delta\sigma}{d\Omega \, dx_1 \, dx_2} = 
\sum_q e_q^2 \left[ h_1^q(x_1, M^2) \, h_1^q(x_2, M^2)
 + h_1^{\bar q}(x_1, M^2) \, h_1^{\bar q}(x_2, M^2) \right] 
\frac{d\Delta\hat{\sigma}}{d\Omega} \>, 
\label{delpp}
\ee
or
\bea
\frac{d\Delta\sigma}{d\varphi \, dM^2 \, dx_F} &=& 
\frac{\alpha^2}{9 \, M^2 s \, (x_1 + x_2)} \> 
\sum_q e_q^2 \left[ h_1^q(x_1, M^2) \, h_1^q(x_2, M^2) \right.
\nonumber \\
& & \hspace{2cm} 
+ \left. h_1^{\bar q}(x_1, M^2) \, h_1^{\bar q}(x_2, M^2) \right] \,
\cos(2\varphi) \> \label{delpp1}
\eea
where $d\Delta\hat{\sigma}/d\Omega$ is given in Eq. (\ref{polqq}) and, again,
all transversity distributions refer to protons.  

Dividing  (\ref{delpp}) by (\ref{unppp}), one gets
\be 
A_{TT}^{p \bar p} \equiv \frac{d\Delta\sigma}{d\sigma} = \hat{a}_{_{TT}} \,
\frac{ \sum_q e_q^2 \left[ h_1^q(x_1, M^2) \, h_1^q(x_2, M^2)
+ h_1^{\bar q}(x_1, M^2) \, h_1^{\bar q}(x_2, M^2) \right] }
{\sum_q e_q^2 \left[ q(x_1, M^2) \, q(x_2, M^2) + \bar q(x_1, M^2) \, 
\bar q(x_2, M^2) \right]}
\label{ATT}
\ee
where $\hat a_{_{TT}}$ is the elementary double spin asymmetry, 
$d\Delta\hat{\sigma}/d\hat{\sigma}$. 
If one detects the polar angle $\theta$ of the lepton pair one has
\be  
\hat a_{_{TT}}(\theta, \varphi) = \frac{\sin^2\theta}{1 + \cos^2\theta} 
\, \cos(2\varphi) \>,
\label{atttp}
\ee
while, when integrating over all polar angles,
\be  
\hat a_{_{TT}}(\varphi) = \frac{1}{2} \, \cos(2\varphi) \>.
\label{attp}
\ee

Before giving numerical estimates of $A_{TT}^{p \bar p}$, 
some comments are in order.
\begin{itemize}
\item
Drell-Yan formulas (\ref{unppp1}) and (\ref{delpp1}) are valid at leading 
order. It is known that NLO contributions to the Drell-Yan 
{\it cross-sections} can be large, especially at typical fixed target 
energies. At these energies, however, NLO corrections to the double 
transverse spin {\it asymmetry} have been found to be moderate \cite{mssv}. 
This makes us confident that the LO accuracy adopted here can give reliable 
results for $A_{TT}$ in the $J/\psi$ region at relatively small values of $s$. 

\item 
Another {\it caveat} concerning Eqs. (\ref{unppp1}), 
(\ref{delpp1}) and (\ref{ATT}) is that they are applicable 
in the continuum region away from resonance thresholds, 
and in particular above the $J/\psi$ and $\psi'$ peak, that is for 
$M \gsim 4$ GeV/$c^2$.

\item
In the kinematical region we wish to explore -- Eq. (\ref{reg}), 
$x_1x_2 \gsim 0.1$ -- sea quark and gluon contributions are 
negligible, hence $A_{TT}^{p \bar p}$ gives direct access to valence 
quark transversity distributions. Actually, taking into account 
the quark charges and the $u$ quark dominance at large $x$, 
Eq. (\ref{ATT}) is essentially given by
\be
\frac{A_{TT}^{p \bar p}}{\hat a_{_{TT}}} \simeq 
\frac{h_1^u(x_1, M^2) \, h_1^u(x_2, M^2)}
{u(x_1, M^2) \, u(x_2, M^2)}  \, \cdot \label{ATTs} 
\ee
which, at $x_1=x_2=\sqrt \tau$, gives 
$[h_1^u(\sqrt \tau, M^2)/u(\sqrt \tau, M^2)]^2$. 
Thus $A_{TT}^{p \bar p}$ represents a unique approach 
to a single transversity distribution, with no flavour 
admixture and no quark-antiquark entanglement. 

\item
Exploring the large $x_1, x_2$ region 
has the clear advantage of offering a direct 
measurement of $h_1^q$; however, it has the disadvantage of limiting such
measurements to a region where, even if $A_{_{TT}}^{p \bar p}$ 
is large, the Drell-Yan 
cross-sections might be too tiny; $q(x_1)$ and $q(x_2)$ in Eq. (\ref{unppp1}) 
are both small at large $x_1, x_2$. 
We shall further discuss this point in the next 
Section.  
\end{itemize} 

In order to give some estimates we have computed the quantity 
\be
\tilde A_{TT}^{p \bar p}(M^2, x_F) \equiv 
\frac{A_{TT}^{p \bar p}}{\hat a_{_{TT}}} \label{ATT1} 
\ee
as given by Eq. (\ref{ATT}), following the procedure of Ref. \cite{bcd}: 
one assumes, as suggested by all relativistic quark model computations
\cite{bdr}, 
\be
h_1^q(x, Q_0^2) = \Delta q(x, Q_0^2) \quad\quad\quad
h_1^{\bar q}(x, Q_0^2) = \Delta \bar q(x, Q_0^2) \label{init}
\ee
at a small scale $Q_0^2$, and then evolves the distributions, according to 
the QCD evolution of $h_1$, 
to the desired scale $M^2$. The initial parton distributions 
are taken from the GRV fits \cite{grv}, which have indeed a small input scale,
$Q_0^2 = 0.23$ (GeV/$c)^2$.

\begin{figure}
\label{fig1}

\hspace{1cm}
\parbox{7cm}{
\scalebox{0.8}{
\includegraphics*[70,440][530,750]{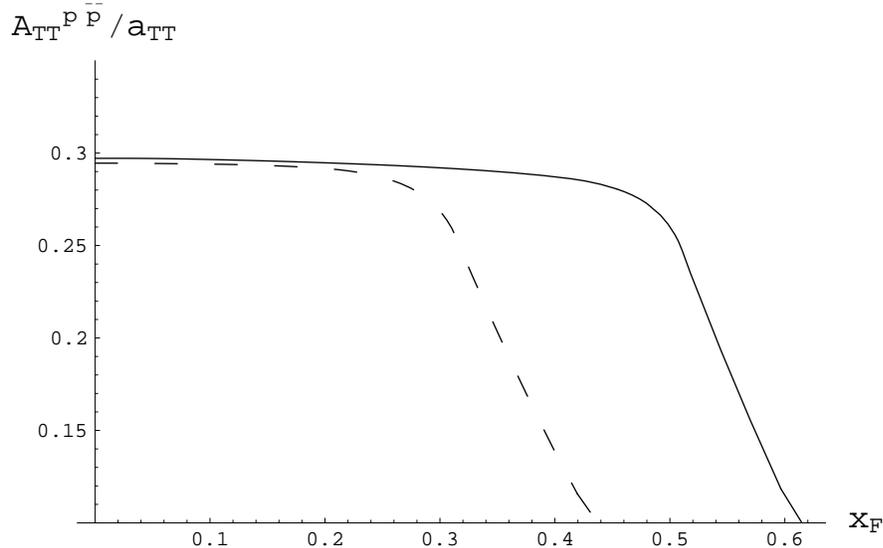}}
}

\parbox{14cm}{
\caption{\small The $p \bar p$ Drell-Yan double transverse spin asymmetry 
$\tilde A_{TT}^{p \bar p}(M^2, x_F)$ as a function of $x_F$, 
for $M = 4$ GeV/$c^2$ (solid curve: $s=45$ GeV$^2$; 
dashed curve: $s = 30$ GeV$^2$).}
}

\end{figure}

The results are shown in Fig. 1 as a function of $x_F$, at $M = 4$ GeV/$c^2$.
The dashed curve corresponds to $s = 30$ GeV$^2$, the solid curve to 
$s = 45$ GeV$^2$. Similar results hold at larger values of $M$.
As one can see, $\tilde A_{TT}^{p \bar p}$ is large 
in the kinematical region considered; notice also the flatness of 
$\tilde A_{TT}^{p \bar p}$  for $x_F \lsim 0.3-0.5$ at fixed $M^2$. 

\vspace{18pt}
\goodbreak
\nd
{\bf 3. {\mbox{\boldmath $A_{TT}$}} for dilepton production via 
{\mbox{\boldmath $J/\psi$}} resonances in 
{\mbox{\boldmath$\pup \!\! \bar p^{\,\uparrow}$}} interactions}
\nobreak
\vspace{6pt}
\nobreak

As we said, the Drell-Yan cross-section might be too small in the 
kinematical region $M \gsim 4$ GeV/$c^2$, $30 \lsim s \lsim 45$ (GeV)$^2$,
which would offer a very good access to valence quark 
transversity distributions. However, it is well known \cite{mmp,sps} that 
the cross-section for dilepton production shows a big bump around $M = 3$ 
GeV/$c^2$, increasing by almost a factor 100 going from $M=4$ GeV/$c^2$ 
to $M=3$ GeV/$c^2$, due to the $J/\psi$ and $\psi'$ resonance production.

The total cross-section for $J/\psi$ production in $p \, \bar p$ 
interactions has been measured to be \cite{sps}
\be
\sigma^{p\bar p \to J/\psi} = (12.0 \pm 5.0) \> {\rm nb}
\quad\quad {\rm at} \quad s = 80 \> ({\rm GeV})^2 \>. \label{jpsics}
\ee
Taking into account a 5.9\% branching ratio for the $J/\psi \to e^-e^+$ 
(or $\mu^-\mu^+$) decay, the value (\ref{jpsics}) is big enough so that,
with a luminosity of the order of $10^{31}$ cm$^{-2}\,$s$^{-1}$, one
expects a number of $p \, \bar p \to J/\psi \to \ell^-\ell^+$ events/year
of the order of $10^5$. $\sigma^{p\bar p \to J/\psi}$ should be approximately
an order of magnitude smaller in the kinematical region (\ref{reg}) discussed 
here; more detailed evaluations can be found in Ref. \cite{pax}. 
These simple estimates also show how the $\ell^-\ell^+$ 
production in the continuum region (probably smaller by almost two orders
of magnitude at $M = 4$ GeV/$c^2$) might be too tiny to allow significant
measurements, unless one could count on very high luminosity machines.

The question is now: how do Eqs. (\ref{unppp1}) and (\ref{ATT}) change
in the $J/\psi$ resonance production region?
From a comparison of the cross-sections measured in $p\,\bar{p}$ and $p\,p$ 
collisions at $s=80$ (GeV)$^2$ \cite{sps} one would conclude that, in the 
energy range we are discussing, the $J/\psi$ production is dominated 
by $q \, \bar q$ fusion \cite{sps}; the dilepton production
in such a resonance region is described in a way analogous to the 
Drell-Yan continuum production, with the elementary cross-section
$q \, \bar q \to \gamma^* \to \ell^-\ell^+$ simply replaced by 
$q \, \bar q \to J/\psi \to \ell^-\ell^+$. As $J/\psi$ is a vector particle, 
like $\gamma^*$, this results in the fact that Eq. (\ref{unppp1})  
applies also to the $p \, \bar p \to J/\psi \to \ell^-\ell^+$  
process with the replacements \cite{lp}:

\be 
16 \pi^2 \alpha^2 e_q^2 \to (g_q^V)^2 \, (g_{\ell}^V)^2 \quad\quad\quad
\frac{1}{M^4} \to \frac{1}{(M^2 - M_{J/\psi}^2)^2 +  
M_{J/\psi}^2  \Gamma_{J/\psi}^2} \> , \label{repl}
\ee
where $g_q^V$ and $g_{\ell}^V$ are the $J/\psi$ vector couplings to 
$q \, \bar q$ and $\ell^-\ell^+$. $\Gamma_{J/\psi}$ is the full width 
of the $J/\psi$ and the new propagator is responsible for the large
observed increase in the cross-section at $M^2 = M_{J/\psi}^2$.

The crucial point is now that, because of the identical helicity
and vector structure of the $\gamma^*$ and $J/\psi$ elementary
channels (all $\gamma^\mu$ couplings) the same replacements hold
for the polarized cross-section, Eqs. (\ref{delpp}, \ref{delpp1}). 
All common factors cancel out in the ratio defining $A_{TT}$, so that one 
has for the $J/\psi$ production region in $p \, \bar p$ processes: 
\be 
A_{TT}^{J/\psi} = \hat{a}_{_{TT}} \,
\frac{ \sum_q \, (g_q^V)^2 \left[ h_1^q(x_1, M^2) \, h_1^q(x_2, M^2)
+ h_1^{\bar q}(x_1, M^2) \, h_1^{\bar q}(x_2, M^2) \right] }
{\sum_q \, (g_q^V)^2 \left[ q(x_1, M^2) \, q(x_2, M^2) + \bar q(x_1, M^2) \, 
\bar q(x_2, M^2) \right]}
\label{ATTjp1}
\ee
with the same $\hat{a}_{_{TT}}$ as given in Eqs. (\ref{atttp}) and
(\ref{attp}).
  
In the large $x_1,x_2$ region we are considering, the $u$ and $d$ valence 
quarks dominate; moreover, we expect the strong $q \, \bar q \, J/\psi$ 
coupling $g_q^V$ to be the same for $u$ and $d$ quarks. Then 
Eq. (\ref{ATTjp1}) further simplifies into: 

\be 
A_{TT}^{J/\psi} \simeq \hat{a}_{_{TT}} \,
\frac{ h_1^u(x_1, M^2) \, h_1^u(x_2, M^2)
+ h_1^d(x_1, M^2) \, h_1^d(x_2, M^2)} 
{u(x_1, M^2) \, u(x_2, M^2) + d(x_1, M^2) \, d(x_2, M^2)} \> \cdot
\label{ATTjp2}
\ee

All models for the transversity distribution agree on having, 
at large $x$, $|h_1^u(x)| \gg |h_1^d(x)|$ \cite{bdr}, so that  
Eq. (\ref{ATTjp1}) simply amounts to  
\be 
 A_{TT}^{J/\psi} \simeq \hat{a}_{_{TT}} \,
\frac{ h_1^u(x_1, M^2) \, h_1^u(x_2, M^2)} 
{u(x_1, M^2) \, u(x_2, M^2)} \> \cdot
\label{ATTjp3}
\ee

Eqs. (\ref{ATTjp1}-\ref{ATTjp3}) are the main new issues 
of this paper; they hold in a region where the unpolarized
cross-section is large and should supply the most direct and viable way
towards measuring transversity distributions. 

In Fig. 2 we show $\tilde A_{TT}^{J/\psi}(x_F) \equiv 
A_{TT}^{J/\psi}/\hat a_{_{TT}}$, as given by Eq. (\ref{ATTjp2})
($M = 3$ GeV/$c^2$), with the same choices for the distribution functions as 
in Fig. 1. We notice that the values of Figs. 1 and 2, that is $A_{TT}$ from 
continuum Drell-Yan production at $M = 4$ GeV/$c^2$ and $A_{TT}$ via $J/\psi$
production at $M = 3$ GeV/$c^2$, are very close.
This should be a well defined test for the validity of our 
Eqs. (\ref{ATTjp1}, \ref{ATTjp2}): in the same region where the 
cross-section shows a change by almost a factor 100, the 
values of $\tilde A_{TT}$ should hardly change.   

One might wonder whether our expressions (\ref{repl}) and (\ref{ATTjp1}), 
which we believe to hold in general at the $J/\psi$ peak, compare with 
existing models and theories for $J/\psi$ production.    
The partonic structure of the asymmetry we derived, that is
$\tilde{A}_{TT}^{J/\psi} \equiv A_{TT}^{J/\psi}/\hat{a_{TT}}$, is quite
independent of the specific mechanism for $J/\psi$ production, provided
this process is dominated by $q \, \bar q$ annihilation, which is the case 
in the kinematic regime we are considering \cite{sps}. As an example of 
a particular mechanism of $J/\psi$ formation, we mention the color evaporation
model \cite{cem}, in which an initial $q \, \bar q$ pair annihilates, via
one-gluon exchange, into a final $c \, \bar c$ pair, which eventually loses
its color by multiple soft gluon emission and hadronizes into a $J/\psi$.
If we assume that the charmonium carries over the polarization of the
$c \, \bar c$ pair, the resulting double spin asymmetry is exactly the one
we obtained above. If, on the contrary, the polarization of the $c \, \bar c$ 
pair is somehow destroyed during the hadronization process, the dilepton 
angular distribution might be different; which means that $\hat{a}_{TT}$ would 
be modified, but nonetheless $\tilde{A}_{TT}^{J/\psi}$ would remain unchanged.
Incidentally, a measurement of the $\theta$-dependence of the cross-section 
in the $q \, \bar q$-dominated regime would shed light on the dynamics of 
$J/\psi$ production.

\begin{figure}[t]
\label{fig2}

\hspace{1cm}
\parbox{7cm}{
\scalebox{0.8}{
\includegraphics*[70,450][530,750]{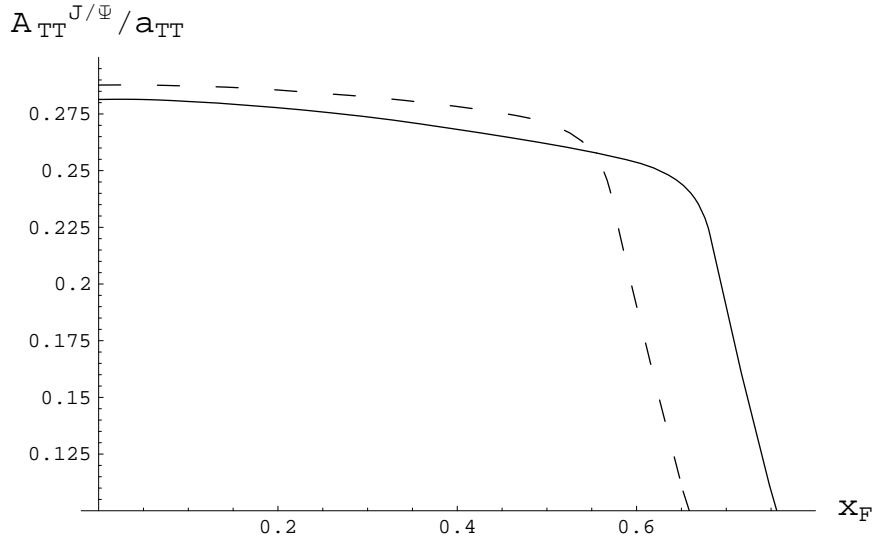}}
}

\parbox{14cm}{
\caption{\small The double transverse spin asymmetry 
$\tilde A_{TT}^{J/\psi}$ for $J/\psi$ production in $p \bar p$ 
collisions, as a function of $x_F$ at $M = 3$ GeV/$c^2$
(solid curve: $s=45$ GeV$^2$; dashed curve: $s = 30$ GeV$^2$).}
}

\end{figure}

\vspace{18pt}
\goodbreak
\nd
{\bf 4. Comments and conclusions}
\nobreak
\vspace{6pt}
\nobreak

The double transverse spin asymmetry $A_{TT}$, in $p \, \bar p$ initiated 
Drell-Yan processes and in kinematical regions exploring the valence quark 
content of the proton, is a unique way of accessing directly the still 
unknown transversity distribution.   

The problem in such a kinematical region might be the smallness of the 
cross-section, which would require very high luminosity beams, difficult 
to achieve with polarized anti-protons. This is the case of the 
continuum region above the $J/\psi$ resonance production,
$M \gsim 4$ GeV/$c^2$, where the cross-section is pQCD calculable. 
However, as we argued above, for the purpose of measuring the transversity 
one could exploit also the data gathered in the $J/\psi$ region, where the
resonance plus continuum cross-section is larger by almost two
orders in magnitude. Specifically, we have shown that
in the $J/\psi$ resonance production region the expression 
of $A_{TT}$ in terms of transversity distributions is essentially the same
as in the continuum case, at least for large $x$ values, Eqs. (\ref{ATTs}) 
and (\ref{ATTjp3}). Our numerical estimates, Figs. 1 and 2, show that $A_{TT}$
is large. This, indeed, offers an experimentally viable and direct
access to $h_1^u(x, M^2)$. Such a measurement could be performed at the
proposed PAX experiment at the GSI-HESR.    
     
\vskip 18pt
\goodbreak
\nd
{\bf Acknowledgements}
\vskip 6pt
We would like to thank the PAX collaboration, which arose our interest in the 
possibility of measuring transversity distributions with polarized protons and 
antiprotons at GSI-HESR.

%\normalsize
%
%\newpage
%
%\noindent {\bf Figure captions}
%
%\vspace{12pt}
%
%\noindent{\bf Fig. 1:\ }
%The kinematical configuration considered in this paper. The $\gamma^*$
%four-momentum defines all our observables; the dependence on the angle
%between the $\gamma^*$-$z$ and the $\gamma^*$-$(\ell^+\ell^-)$ planes is
%integrated over in Eq.s~(\ref{ddy2}) and (\ref{udy2}).

%\vspace{8pt}

\end{document}